\documentclass[12pt]{article}
\usepackage{a4wide}
\usepackage{epsfig}

\newcommand{\pfrac}[2]{\left(\frac{#1}{#2}\right)}
\newcommand{\eps}{\varepsilon}
\newcommand{\Li}{\mathop{\rm Li}\nolimits}
\newcommand{\Cl}{\mathop{\rm Cl}\nolimits}
\newcommand{\ELi}{\mathop{\rm ELi}\nolimits}

\begin{document}
\thispagestyle{empty} 
\begin{flushright}
MITP/18-032
\end{flushright}

\begin{center}
{\Large\bf Coordinate space calculation of two- and three-loop sunrise-type
diagrams, elliptic functions and truncated Bessel integral identities}

\vspace{1truecm}
{\large \bf S.~Groote$^1$ and J.G.~K\"orner$^2$}\\[.4truecm]
$^1$F\"u\"usika Instituut, Tartu \"Ulikool,
  W.~Ostwaldi 1, EE-50411 Tartu, Estonia\\[.3truecm]
$^2$Institut f\"ur Physik der Johannes-Gutenberg-Universit\"at,\\
  Staudinger Weg 7, D-55099 Mainz, Germany
\end{center}

\begin{abstract}
We integrate three-loop sunrise-type vacuum diagrams in $D_0=4$
dimensions with four different masses using configuration space techniques.
The finite parts of our results are in numerical agreement with corresponding
three-loop calculations in momentum space. Using some of the closed form
results of the momentum space calculation we arrive at new integral identities
involving truncated integrals of products of Bessel functions. For the
non-degenerate finite two-loop sunrise-type vacuum diagram in $D_0=2$
dimensions we make use of the known closed form $p$-space result to express
the moment of a product of three Bessel functions in terms of a sum of
Clausen polylogarithms. Using results for the nondegenerate two-loop sunrise
diagram from the literature in $D_0=2$ dimensions we obtain a Bessel function
integral identity in terms of elliptic functions.
\end{abstract}

\newpage

\section{Introduction}
The computation of higher order vacuum diagrams in quantum field theory has
been extended to ever higher loop levels involving different degrees of
mass degeneracy and/or zero mass values for their mass configurations.
The multi-loop calculations have
traditionally been carried out in momentum space ($p$-space). For the
sunrise-type subclass of these higher order vacuum diagrams it is much simpler
to integrate the diagrams using configuration-space ($x$-space) techniques.
The comparison of the results of the two calculations will provide a welcome
nontrivial cross check on the correctness of the respective $p$- and $x$-space
calculations in as much as a subclass of the vacuum diagrams are of
sunrise-type. In addition, if one can avail of closed form $p$-space results,
one arrives at new nontrivial integral identities involving moments of Bessel
functions. In the case of the two-loop nondegenerate sunrise diagram this
leads to an identity of moments of Bessel functions in terms of elliptic
integrals. We mention that Bessel function integral identities involving
elliptic functions have been studied before in
Refs.~\cite{Bailey:2008ib,Broadhurst:2008mx}.

This paper was triggered by the recent appearance of two $p$-space
calculations of the nondegenerate three-loop vacuum diagrams (all four masses
different). We numerically confirm the results of the vacuum sunrise-type
diagrams in Refs.~\cite{Freitas:2016zmy,Martin:2016bgz} using $x$-space
techniques. In Sec.~2 we briefly review some general features of the $x$-space
approach and describe the splitting technique used to separate singular
(analytical) and finite (numerical) parts in $D=D_0-2\eps$ spacetime
dimensions where $D_0=4$ in the present application. The particular form of
the splitting technique
preserves the inherent symmetry of the nondegenerate multiloop vacuum and
sunrise integrals w.r.t.\ the exchange of different rungs or masses in the
diagrams. This leads to the notion of truncated Bessel integrals. In Sec.~3 we
use closed form results for the finite parts of the three-loop vacuum diagrams
from the literature to obtain new integral identities for a set of truncated
Bessel integrals. In Sec.~4 we compare results for the nondegenerate two-loop
sunrise diagram with results from Refs.~\cite{Adams:2014vja,Adams:2015pya,%
Remiddi:2016gno,Primo:2017ipr} for $D_0=2$ to obtain again an integral
identity. Our conclusions are given in Sec.~5, and series expansions for the
Bessel functions are found in the Appendix.

\section{Splitting technique for Bessel integrals}
The configuration space calculation of $n$-loop sunrise-type diagrams with
$N=n+1$ different masses in arbitrary spacetime dimensions has been studied
by us in a series of papers~\cite{Groote:1998ic,Groote:1998wy,Groote:1999cx,%
Groote:1999cn,Groote:2000kz,Groote:2004qq,Groote:2005ay,Groote:2012pa}. The
starting point is the central identity for the $p$-space correlator function
$\tilde\Pi(p;m_1,\ldots,m_{n+1})$ given by
\begin{equation}\label{besint}
\tilde\Pi(p;m_1,\ldots,m_{n+1})=2\pi^{\lambda+1}\int_0^\infty
\pfrac{px}2^{-\lambda}J_\lambda(px)\prod_{i=1}^{n+1}D(x,m_i)x^{2\lambda+1}dx,
\end{equation}
where $p=\sqrt{p^2}$ and $x=\sqrt{x^2}$ are the absolute values of the
four-vectors of four-momentum and spacetime, respectively. The relation of the
spacetime dimension to the parameter $\lambda$ is given by $D=2\lambda+2$.
$n$-loop sunrise-type integrals are UV divergent for $nD\ge 2(n+1)$ (or
$n\lambda>1$). In order to parametrize the singularities, we use dimensional
regularization by choosing $D=D_0-2\eps$. $J_\lambda(z)$ is the Bessel
function of the first kind, and
\begin{equation}
D(x,m)=\int\frac{d^Dp}{(2\pi)^D}\frac{e^{ip_\mu x^\mu}}{p^2+m^2}
  =\frac{(mx)^\lambda K_\lambda(mx)}{(2\pi)^{\lambda+1}x^{2\lambda}}
\end{equation}
is the free propagator in (Euclidean) spacetime for a particle line with mass
$m$. The function $K_\lambda(mx)$ denotes the Bessel function of the second
kind, also known as the McDonald function. The series expansions for both
Bessel functions are found in the Appendix. In the case that the particle mass
vanishes ($m=0$), the propagator simplifies to
\begin{equation}
D(x,0)=\int\frac{d^Dp}{(2\pi)^D}\frac{e^{ip_\mu x^\mu}}{p^2}
  =\frac{\Gamma(\lambda)}{4\pi^{\lambda+1}x^{2\lambda}},
\end{equation} 
since the
corresponding Bessel function of the second kind is replaced by a simple power
dependence. A corresponding simplification occurs in the case of a vacuum
diagram ($p^2\to 0$) where one has
\begin{equation}
\pfrac{px}2^{-\lambda}J_\lambda(px)\to\frac1{\Gamma(\lambda+1)}.
\end{equation}
Additional features can easily be implemented in the configuration space
calculus. For instance, higher powers of propagators can be inserted by
calculating the derivative with respect to the corresponding squared mass,
i.e.\
\begin{equation}
\tilde D^{(\kappa)}(p,m)=\frac1{(p^2+m^2)^{\kappa+1}}=\frac1{\Gamma(\kappa+1)}
  \pfrac{-\partial}{\partial m^2}^\kappa\frac1{p^2+m^2}
\end{equation}
which leads to
\begin{equation}
D^{(\kappa)}(x,m)=\int\frac{d^Dp}{(2\pi)^D}
  \frac{e^{ip_\mu x^\mu}}{(p^2+m^2)^{\kappa+1}}
  =\frac{(m/x)^{\lambda-\kappa}}{(2\pi)^{\lambda+1}2^\kappa\Gamma(\kappa+1)}
  K_{\lambda-\kappa}(mx).
\end{equation}
Bessel integral identities involving up to three Bessel functions have been
compiled in integral tables (cf.\ for instance
Refs.~\cite{Prudnikov,Gradshteyn}).
Some Bessel function integrals involving up to six Bessel functions can be
found in Refs.~\cite{Bailey:2008ib,Broadhurst:2008mx}. 
To proceed one needs
analytical expressions for the singular part of the
result. However, this problem can be circumvented because in certain
kinematical cases
(small momentum or specific mass configurations) one can expand the Bessel
functions in Taylor series to obtain a series expansion of the Bessel integral.
In Ref.~\cite{Groote:2005ay} we explained how the Bessel integral
can be calculated by expanding the Bessel function of the first kind or, in
case of vacuum diagrams, one or more McDonald functions.\footnote{In this
approach, at least one single McDonald function has to stay unexpanded to
guarantee the convergence of the integrand for large values of $x$.} However,
in such an expansion the symmetry with respect to the exchanges of the
different masses in the diagram is no longer manifest.

In order to avoid this problem, we use another approach to isolate the
divergent parts of the integrals. Namely, we use the Gaussian factor
$e^{-\mu^2x^2}$ to protect the integrand against the divergent upper limit.
The mass parameter $\mu$ in the exponent defines a new regularization scale
parameter. In detail, we add and subtract a series expansion of the product of
Bessel functions multiplied by $e^{\mu^2x^2}$ to a specified order. The order
of the series expansion is chosen such that the difference is no longer
singular at $x=0$. The series can be integrated analytically using
\begin{equation}
\int_0^\infty x^{p-1}e^{-\mu^2x^2}dx=\frac12\mu^{-p}\Gamma(p/2)
\end{equation}
where the r.h.s. can be expanded in $\eps$. The
integral
\begin{equation}
\int_0^\infty[f(x)]_xdx
\end{equation}
will be called truncated integral in the following. The truncated integral
involves a series truncation
\begin{equation}
  \label{trunc}
[f(x)]_x:=f(x)-e^{-\mu^2x^2}\left(e^{\mu^2x^2}f(x)+O(x)\right),
\end{equation}
where $f(x)$ is the original integrand in Eq.~(\ref{besint}). The difference
is integrable numerically even in the
limit $\eps\to 0$ and constitutes a part of the finite contribution. Of
course, the regularized integral depends on the (arbitrary) mass scale $\mu$.
However, taking into account the integrated series expansion leading to a
power series in inverse powers of $\eps$, the scale dependence cancels out, as
it should be according to the construction.

With the techniques described in this section we have recalculated the
nondegenerate three-loop sunrise-type vacuum diagrams in
Refs.~\cite{Freitas:2016zmy,Martin:2016bgz} and 
have checked our results against the results given in
Refs.~\cite{Freitas:2016zmy,Martin:2016bgz}. For the relations
\begin{eqnarray}
\tilde\Pi(0;m_1,m_2,m_3,m_4)&=&-\frac{e^{-3\gamma_E\eps}}{4^{6-3\eps}}
  M(1,1,1,1,0,0;m_1^2,m_2^2,m_3^2,m_4^2,0,0),\nonumber\\
\tilde\Pi(0;\dot m_1,m_2,m_3,m_4)&=&-\frac{e^{-3\gamma_E\eps}}{4^{6-3\eps}}
  M(2,1,1,1,0,0;m_1^2,m_2^2,m_3^2,m_4^2,0,0)
\end{eqnarray}
(cf.\ Ref.~\cite{Freitas:2016zmy}, a dot at the mass entry in $\tilde\Pi$
indicates the multiplicity) and
\begin{eqnarray}
\tilde\Pi(0;m_1,m_2,m_3,m_4)&=&\frac{e^{-3\gamma_E\eps}}{(4\pi)^{6-3\eps}}
  {\bf E}(m_1^2,m_2^2,m_3^2,m_4^2),\nonumber\\
\tilde\Pi(0;\dot m_1,m_2,m_3,m_4)&=&\frac{e^{-3\gamma_E\eps}}{(4\pi)^{6-3\eps}}
  {\bf F}(m_1^2,m_2^2,m_3^2,m_4^2)
\end{eqnarray}
(cf.\ possible mass constellations in Ref.~\cite{Martin:2016bgz}) we have
found numerical agreement. In case of Ref.~\cite{Freitas:2016zmy}, this
comparison was performed with the help of the code
TVID~\cite{Bauberger:2017nct} for randomly generated sets of four different
masses in a total sample of 1000 mass configurations.  

For the genuine sunrise diagram, a comparison is possible also to results with
equal~\cite{Bloch:2013tra} and different masses~\cite{Broedel:2017siw}
expressed in terms of multiple elliptic polylogarithms~\cite{Brown:2011}.

\begin{figure}\begin{center}
\epsfig{figure=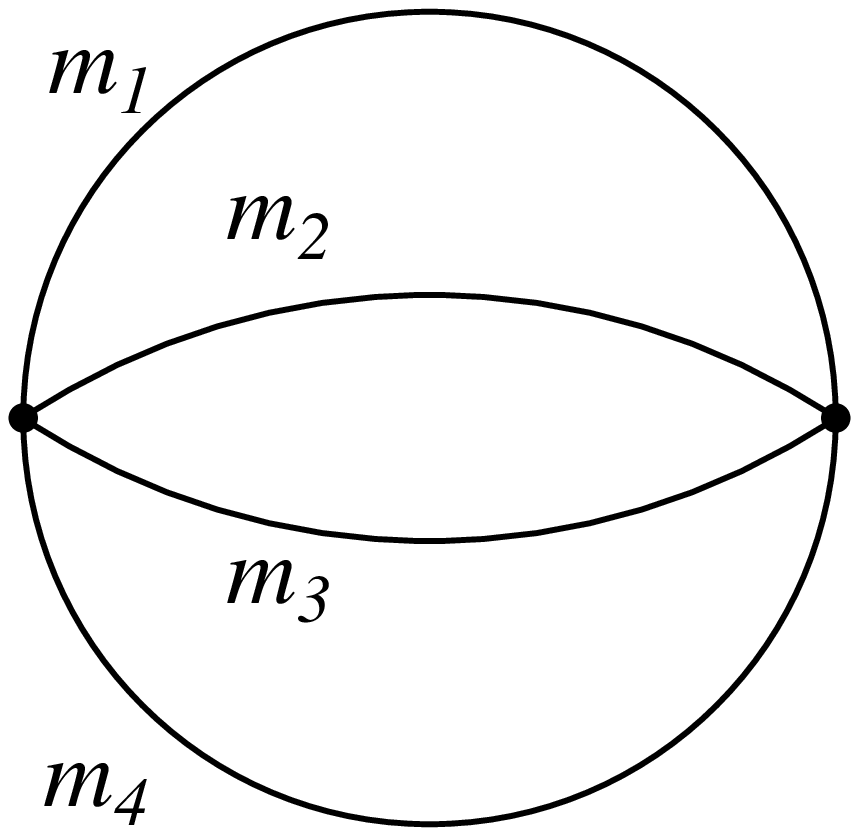, scale=0.4}\qquad
\epsfig{figure=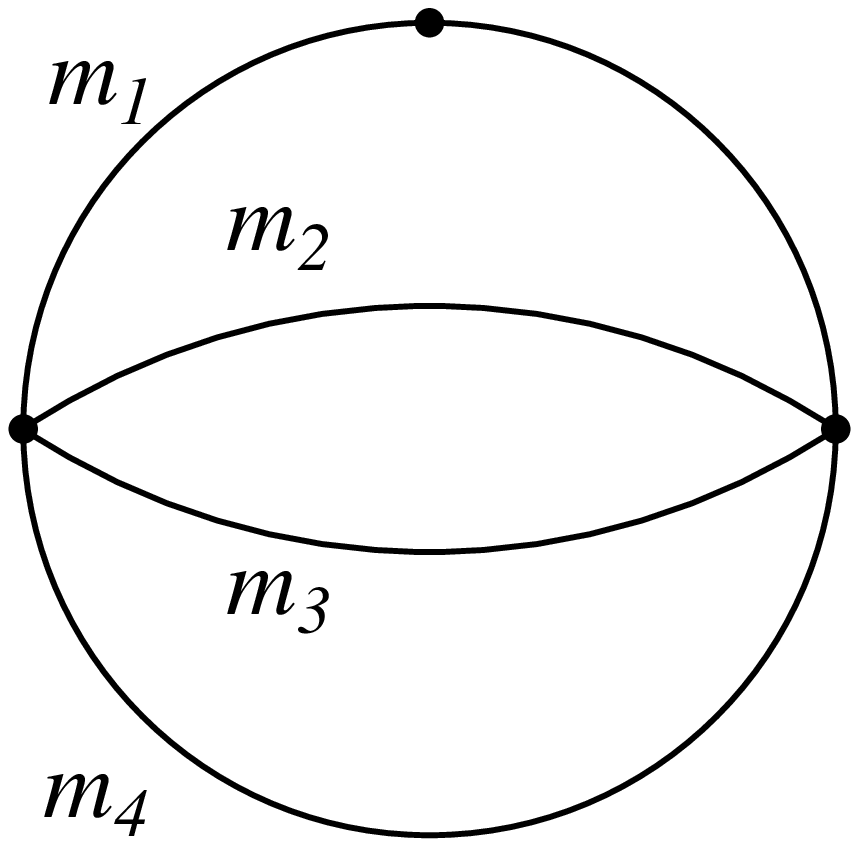, scale=0.4}\qquad
\epsfig{figure=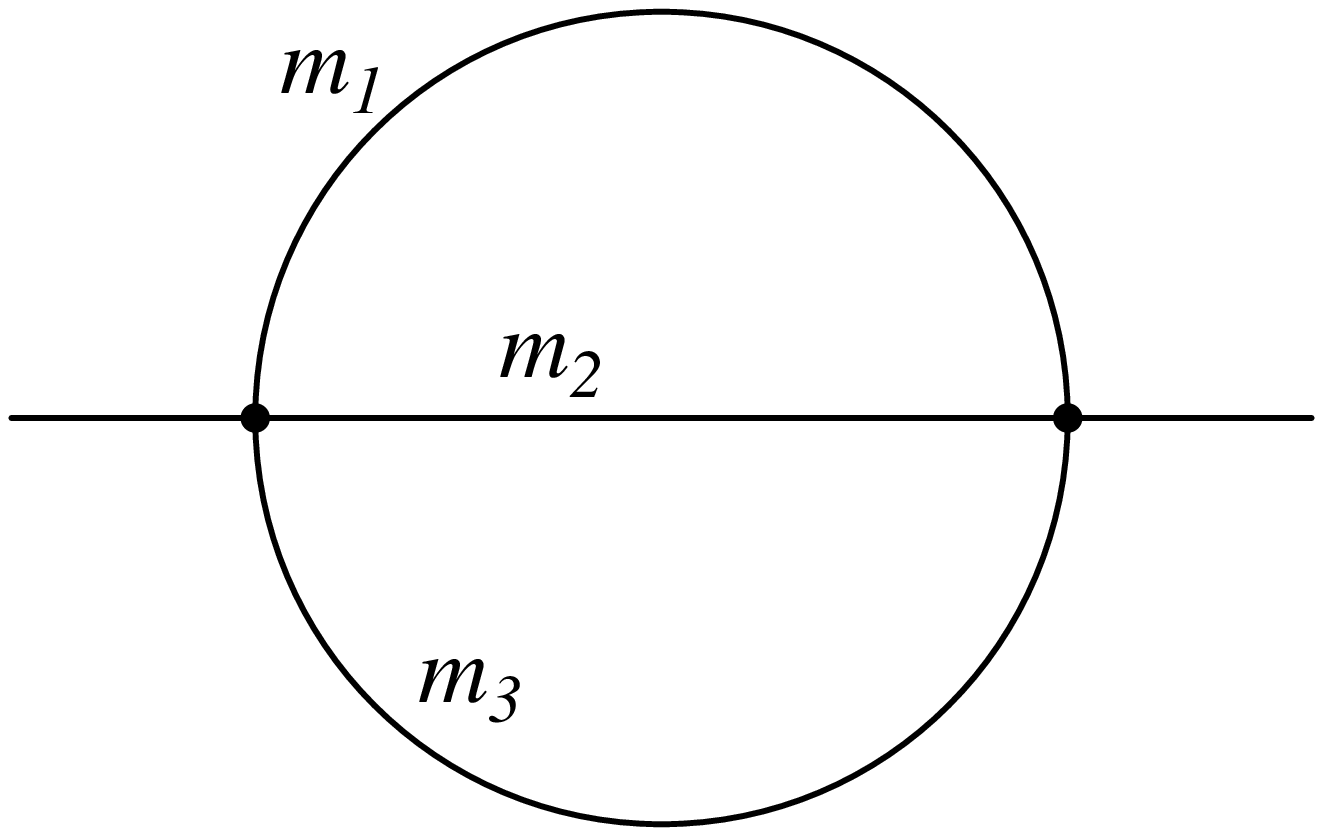, scale=0.4}\\
(a)\kern106pt(b)\kern132pt(c)\kern24pt\strut
\caption{\label{diagrams} Three-loop nondegenerate vacuum diagrams (a) without
  and (b) with a derivative indicated by a dot on the top line. Diagram (c)
  shows a two-loop (genuine) sunrise diagram with three different masses}
\end{center}\end{figure}

\section{Truncated Bessel integrals for $D_0=4$}
For some sunrise-type diagrams, $p$-space calculations have led to analytic
closed form results. These results can be used as input to derive Bessel
function integral identities by comparison with the corresponding $x$-space
results. As emphasized before, sunrise-type diagrams in $D_0=4$ dimensions
have to be regularized, implying that the corresponding integral identities
have to be formulated in terms of truncated Bessel integrals. A complete list
of analytical results for vacuum diagrams of sunrise-type topology known up to
three-loop order is given in Ref.~\cite{Martin:2016bgz}. We have used the
analytical results for ${\bf E}(u,v,y,z)$ (corresponding to
Fig.~\ref{diagrams}(a)) and ${\bf F}(u,v,y,z)$ (corresponding to
Fig.~\ref{diagrams}(b)) given in Ref.~\cite{Martin:2016bgz} to calculate
truncated Bessel integrals. By choosing $\mu=1$ we obtain
\begin{eqnarray}
\lefteqn{\int_0^\infty\left[\frac{32aK_1(ax)}{x^4}\right]_xdx
  \ =\ 4-\frac{17}8a^4+\frac12a^2(16+5a^2)\ell_a-a^4\ell_a^2
  -\frac18a^4\pi^2,}\nonumber\\[12pt]
\lefteqn{\int_0^\infty\left[\frac{32a^2K_1(ax)^2}{x^3}\right]_xdx
  \ =\ 4-\frac{19}4a^4+a^2(16+3a^2)\ell_a+2a^4\ell_a^2-\frac83a^4\ell_a^3
  \strut}\nonumber\\&&\kern4truecm\strut
  +\frac1{12}a^4\pi^2-\frac13a^4\pi^2\ell_a+\frac{10}3a^4\zeta(3),
  \nonumber\\[12pt]
\lefteqn{\int_0^\infty\left[\frac{32a^3K_1(ax)^3}{x^2}\right]_xdx
  \ =\ 4-\frac{63}8a^4+\frac32a^2(16+a^2)\ell_a+9a^4\ell_a^2-8a^4\ell_a^3
  \strut}\nonumber\\&&\kern4truecm\strut
  -\frac58a^4\pi^2-a^4\pi^2\ell_a+\frac32a^4\psi'\pfrac13+2a^4\zeta(3),
  \nonumber\\[12pt]
\lefteqn{\int_0^\infty\left[\frac{32a^4K_1(ax)^4}x\right]_xdx
  \ =\ 4-\frac{23}2a^4+2a^2(16-a^2)\ell_a+20a^4\ell_a^2-16a^4\ell_a^3
  \strut}\nonumber\\&&\kern4truecm\strut
  +\frac56a^4\pi^2-2a^4\pi^2\ell_a+4a^4\zeta(3),\nonumber\\[12pt]
\lefteqn{\int_0^\infty\left[\frac{32abK_1(ax)K_1(bx)}{x^3}\right]_xdx
  \ =\ 4-\frac12a^2b^2-\frac{17}8(a^2+b^2)-\frac1{24}(a^4-4a^2b^2+b^4)\pi^2
  \strut}\nonumber\\&&\strut
  +\frac{a^2}2(16+5a^2-2b^2)\ell_a+\frac{b^2}2(16-2a^2+5b^2)\ell_b
  -\frac16a^2b^2\pi^2(\ell_a+\ell_b)\strut\nonumber\\&&\strut
  -\frac12(a^4-b^4)\ell_a^2-(a^4-4a^2b^2+b^4)\ell_a\ell_b
  +\frac12(a^4-b^4)\ell_b^2\strut\nonumber\\&&\strut
  +\frac23a^2b^2\ell_a^3-2a^2b^2\ell_a^2\ell_b-2a^2b^2\ell_a\ell_b^2
  +\frac23a^2b^2\ell_b^3+\frac43a^2b^2\zeta(3)\strut\nonumber\\&&\strut
  +\frac14(a^4-b^4)\left[\Li_2\left(1-\frac{a^2}{b^2}\right)
  -\Li_2\left(1-\frac{b^2}{a^2}\right)\right]\strut\nonumber\\&&\strut
  +a^2b^2\left[\Li_3\pfrac{a^2}{b^2}+\Li_3\pfrac{b^2}{a^2}
  -\frac12\ln\pfrac{a^2}{b^2}\left\{\Li_2\pfrac{a^2}{b^2}
  -\Li_2\pfrac{b^2}{a^2}\right\}\right],\nonumber\\[12pt]
\lefteqn{\int_0^\infty\left[\frac{32a^2b^2K_1(ax)^2K_1(bx)^2}x\right]_xdx
  \ =\ 4-2a^2b^2-\frac{19}4(a^4+b^4)+\frac1{12}(a^4+8a^2b^2+b^4)\pi^2
  \strut}\nonumber\\&&\strut
  +a^2(16+3a^2-4b^2)\ell_a+b^2(16-4a^2+3b^2)\ell_b
  -\frac{a^2}3(a^2+2b^2)\pi^2\ell_a-\frac{b^2}3(2a^2+b^2)\pi^2\ell_b
  \strut\nonumber\\&&\strut
  +2a^2(a^2-2b^2)\ell_a^2+24a^2b^2\ell_a\ell_b-2b^2(2a^2-b^2)\ell_b^2
  +\frac23(5a^4-4a^2b^2+5b^4)\zeta(3)\strut\nonumber\\&&\strut
  -\frac43(a^2-b^2)^2\ell_a^3-4(a^4+2a^2b^2-b^4)\ell_a^2\ell_b
  +4(a^4-2a^2b^2-b^4)\ell_a\ell_b^2-\frac43(a^2-b^2)^2\ell_b^3
  \strut\nonumber\\&&\strut
  +(a^2-b^2)^2\Bigg[\Li_3\pfrac{a^2}{b^2}+\Li_3\pfrac{b^2}{a^2}
  \strut\nonumber\\&&\strut\qquad
  -\frac12\ln\pfrac{a^2}{b^2}\left\{\Li_2\pfrac{a^2}{b^2}
  -\Li_2\left(1-\frac{a^2}{b^2}\right)-\Li_2\pfrac{b^2}{a^2}
  +\Li_2\left(1-\frac{b^2}{a^2}\right)\right\}\Bigg],\nonumber\\[12pt]
\lefteqn{\int_0^\infty\left[\frac{32ab^2K_1(ax)K_1(bx)^2}{x^2}\right]_xdx
  \ =\ 4-a^2b^2-\frac18(17a^4+38b^4)-\frac1{24}(a^4-8a^2b^2-2b^4)\pi^2
  \strut}\nonumber\\&&\strut
  +\frac12a^2(16+5a^2-4b^2)\ell_a+b^2(16-2a^2+3b^2)\ell_b
  -\frac13a^2b^2\pi^2\ell_a-\frac13b^2(a^2+b^2)\pi^2\ell_b
  \strut\nonumber\\&&\strut
  -2a^2(a^2-4b^2)\ell_a\ell_b+(a^4+2b^4)\ell_b^2
  -8a^2b^2\ell_a\ell_b^2+\frac83b^2(a^2-b^2)\ell_b^3
  +\frac23b^2(4a^2-b^2)\zeta(3)\strut\nonumber\\&&\strut
  -\frac a2(a^2+2b^2)\sqrt{a^2-4b^2}\left[
  \Li_2\left(-\frac{a-\sqrt{a^2-4b^2}}{a+\sqrt{a^2-4b^2}}\right)
  -\Li_2\left(-\frac{a+\sqrt{a^2-4b^2}}{a-\sqrt{a^2-4b^2}}\right)\right]
  \strut\nonumber\\&&\strut
  -2b^2(a^2-b^2)\left[\Li_3\left(-\frac{a-\sqrt{a^2-4b^2}}{a+\sqrt{a^2-4b^2}}
  \right)+\Li_3\left(-\frac{a+\sqrt{a^2-4b^2}}{a-\sqrt{a^2-4b^2}}\right)
  \right],
\end{eqnarray}
where
\begin{equation}
\ell_a:=\ln\pfrac a2+\frac12\gamma_E,\qquad
\ell_b:=\ln\pfrac b2+\frac12\gamma_E.
\end{equation}
The parameters $a$ and $b$ are mass parameters associated with the vacuum
diagrams. All seven of these truncated Bessel function integral identities
have been checked numerically using MATHEMATICA.

\section{Bessel integrals and elliptic functions for $D_0=2$}
It is well known that the correlator function
$\tilde\Pi(p;m_1,\ldots,m_{n+1})$ for the sunrise-type diagrams is finite in
$D_0=2$ dimensions. Therefore, the corresponding Bessel integrals need not to
be truncated.

For the vacuum integrals ($p^2=0$) the comparison with results from
Refs.~\cite{Usyukina:1992jd,Lu:1992ny,Bern:1996ka,Adams:2013kgc} results in
\begin{equation}
\int_0^\infty2xK_0(m_1x)K_0(m_2x)K_0(m_3x)dx=\frac1{\sqrt{-\lambda}}
  \left(\Cl_2(\alpha_1)+\Cl_2(\alpha_2)+\Cl_2(\alpha_3)\right),
\end{equation}
where $\lambda=\lambda(m_1^2,m_2^2,m_3^2)
=m_1^4+m_2^4+m_3^4-2m_1^2m_2^2-2m_1^2m_3^2-2m_2^2m_3^2$ is the K\"all\'en
function, and $\Cl_p(z)$ is the Clausen polylogarithm,
\begin{equation}
\Cl_p(\alpha):=\frac1{2i}\left(\Li_p(e^{i\alpha})-\Li_p(e^{-i\alpha})\right)
\end{equation}
with arguments $\alpha_i=2\arctan(\sqrt{-\lambda}/\delta_i)$, where
\begin{equation}
\delta_1=-m_1^2+m_2^2+m_3^2,\qquad
\delta_2=m_1^2-m_2^2+m_3^2,\qquad
\delta_3=m_1^2+m_2^2-m_3^2.
\end{equation}

For the more general case $p^2\ne 0$ (cf.\ Fig.~\ref{diagrams}(c)), analytical
results have been given in Refs.~\cite{Adams:2014vja,Adams:2015pya} in terms
of elliptic polylogarithms
\begin{equation}
\ELi_{m;n}(x;y;z)=\sum_{j=1}^\infty\sum_{k=1}^\infty\frac{x^j}{j^m}
  \frac{y^k}{k^n}z^{jk}.
\end{equation}
In order to understand the construction, we start with the complete
elliptic integral
\begin{equation}
K(k)=\int_0^1\frac{dt}{\sqrt{(1-t^2)(1-k^2t^2)}}
\end{equation}
for the arguments
\begin{equation}
k_\pm:=\sqrt{\frac12\pm\frac{p^4+2(m_1^2+m_2^2+m_3^2)p^2
  +\lambda(m_1^2,m_2^2,m_3^2)}{2\sqrt{(p^2+\mu_1^2)(p^2+\mu_2^2)(p^2+\mu_3^2)
  (p^2+\mu_4^2)}}},
\end{equation}
where $\mu_i$ are the (pseudo)thresholds
\begin{equation}
\mu_1=m_1+m_2-m_3,\quad
\mu_2=m_1-m_2+m_3,\quad
\mu_3=-m_1+m_2+m_3,\quad
\mu_4=m_1+m_2+m_3.
\end{equation}
While the second argument of the elliptic polylogarithm is $-1$, the last
argument is given by $-q$ where
\begin{equation}
q=\exp\left(-\pi\frac{K(k_+)}{K(k_-)}\right).
\end{equation}
Defining a modified Clausen dilogarithm by
\begin{equation}
\tilde{\Cl}_2(\tilde\alpha):=\frac1{2i}\left(\Li_2(e^{i\tilde\alpha})
  -\Li_2(e^{-i\tilde\alpha})+2\ELi_{2;0}(e^{i\tilde\alpha};-1;-q)
  -2\ELi_{2;0}(e^{-i\tilde\alpha};-1;-q)\right),
\end{equation}
the comparison gives
\begin{equation}\label{JKKK}
\int_0^\infty 2xJ_0(px)K_0(m_1x)K_0(m_2x)K_0(m_3x)dx
  =\frac{2K(k_+)\left(\tilde{\Cl}_2(\tilde\alpha_1)
  +\tilde{\Cl}_2(\tilde\alpha_2)+\tilde{\Cl}_2(\tilde\alpha_3)\right)}{
  {}^4\sqrt{(p^2+\mu_1^2)(p^2+\mu_2^2)(p^2+\mu_3^2)(p^2+\mu_4^2)}\,\pi},
\end{equation}
where
\begin{equation}
\tilde\alpha_i=\pi\frac{F(k_i^{-1},k_+)}{K(k+)},\qquad
k_i=\sqrt{k_+^2+\frac{4m_j^2m_k^2}{\sqrt{(p^2+\mu_1^2)(p^2+\mu_2^2)
  (p^2+\mu_3^2)(p^2+\mu_4^2)}}}
\end{equation}
($(i,j,k)$ is a cyclic permutation of $(1,2,3)$) and the incomplete elliptic
integral is given by
\begin{equation}
F(t_0,k):=\int_0^{t_0}\frac{dt}{\sqrt{(1-t^2)(1-k^2t^2)}}.
\end{equation}

The result~(\ref{JKKK}) holds only for the component of the Riemann sheet
connected to the equal mass case $m_1=m_2=m_3$. Problems for other mass
configurations with Eq.~(\ref{JKKK}) can be revealed already in the limit
$p^2\to 0$. Obviously, in his limit the main square root simplifies to
\begin{equation}
\sqrt{(p^2+\mu_1^2)(p^2+\mu_2^2)(p^2+\mu_3^2)(p^2+\mu_4^2)}\to-\lambda.
\end{equation}
Further, one has $k_-=1$ and $k_+=0$. Thus $q\to 0$, and the elliptic
polylogarithms vanish.
However, for the arguments of the (modified) Clausen dilogarithms one obtains
\begin{equation}
\tilde\alpha_i=2\arcsin\left(\sqrt{\frac{-\lambda}{4m_j^2m_k^2}}\right)
  \quad\Leftrightarrow\quad
\frac{-\lambda}{4m_j^2m_k^2}=\sin^2\pfrac{\tilde\alpha_i}2,
\end{equation}
leading to
\begin{equation}
\tan\pfrac{\tilde\alpha_i}2=\frac{\sqrt{-\lambda}}{\sqrt{\delta_i^2}}
  \qquad\Leftrightarrow\qquad
\tilde\alpha_i=2\arctan\pfrac{\sqrt{-\lambda}}{\sqrt{\delta_i^2}}.
\end{equation}
Only in the case $\delta_i>0$  ($i=1,2,3$) and if one is close to the equal mass
case, one can reconstruct the $\alpha_i$. As mentioned in
Ref.~\cite{Bogner:2017vim}, the problem is caused by the standard convention
for mathematical software which is in conflict with the Feynman prescription
used in quantum field theory. In Ref.~\cite{Kaldamae:2014fua} the region in
the projective plane restricted by $\lambda(m_1^2,m_2^2,m_3^2)<0$ is called
the K\"all\'en triangle. In terms of Ref.~\cite{Kaldamae:2014fua} the
restriction $\delta_i>0$ ($i=1,2,3$) leads to a region enclosed by the three
circles of the projective plane about the three corners through the midpoints
of the edges, deforming the K\"all\'en triangle to a hyperbolic
triangle.\footnote{Because of a definition of projectivity in squared masses
instead of a definition in linear masses, the
``K\"all\'en triangle'' is deformed to a circle in Ref.~\cite{Burda:2017tcu}, while the region $\delta_i>0$
would be given by a triangle.}
We emphasize that our configuration space method is free from such kind of
problems and, therefore, can be used to check existing analytical results in
these problematic cases.

\section{Summary and conclusions}
We have recalculated nondegenerate sunrise-type three-loop vacuum integrals in
$x$-space and have found numerical agreement with corresponding results
obtained by $p$-space calculations. The agreement of the respective results
can be considered to be a necessary -- but not sufficient -- check on the
rather involved $p$-space three-loop vacuum diagram calculations. It is not
difficult to extend the $x$-space calculation of the sunrise-type vacuum
diagrams to higher loop orders where they can again be used to check on the
results of higher loop vacuum diagram calculations in $p$-space when they
become available in the future. In fact, the authors of
Ref.~\cite{Luthe:2016sya} have checked their $p$-space results on the
degenerate (all masses equal) five-loop sunrise-type vacuum integrals against
the $x$-space results of e.g.\ Ref.~\cite{Groote:2005ay}.

By comparing the results of some $p$-space and $x$-space calculations of
specific two- and three-loop sunrise diagrams we have presented a number of
novel Bessel function integral identities that would be difficult to prove
otherwise. For divergent sunrise integrals we have introduced a regularization
procedure in terms of a truncation which lead to a number of novel truncated
Bessel function integral identities. For the nondegenerate genuine two-loop
sunrise
diagram in two dimensions we have obtained, again by comparing $x$- and
$p$-space results, a Bessel function integral identity in terms of elliptic
functions. It appears that the techniques discussed in this paper can open a
Pandora's box of new Bessel function integral identities.

\subsection*{Acknowledgements}
We would like to thank A.~Freitas for very constructive e-mail exchanges on
the mutual numerical evaluation of the vacuum integrals.
This research was supported by the Estonian Research Council under Grant
No.~IUT2-27. S.G.\ acknowledges the support of the theory group THEP at the
Institute of Physics and of the Cluster of Excellence PRISMA at the University
of Mainz. 

\begin{appendix}

\section{Series expansions of the Bessel functions}
\setcounter{equation}{0}\def\theequation{A\arabic{equation}}
The two Bessel functions $J_\lambda(z)$ and $K_\lambda(z)$ used in this paper
are defined via the following Taylor series expansions,
\begin{eqnarray}
\lefteqn{J_\lambda(z)\ =\ \pfrac{z}2^\lambda\sum_{k=0}^\infty
  \frac{(-z^2/4)^k}{k!\Gamma(\lambda+k+1)}}\nonumber\\
  &=&\frac{(z/2)^\lambda}{\Gamma(1+\lambda)}
  \Bigg(1-\frac{(z/2)^2}{(1+\lambda)}+\frac{(z/2)^4}{2(1+\lambda)(2+\lambda)}
  -\frac{(z/2)^6}{6(1+\lambda)(2+\lambda)(3+\lambda)}+O(z^8)\Bigg),
  \nonumber\\\\
\lefteqn{K_\lambda(z)\ =\ \frac\pi{2\sin(\lambda\pi)}
  \left(I_{-\lambda}(z)-I_\lambda(z)\right)\ =\ \frac12\Gamma(\lambda)
  \Gamma(1-\lambda)\left(I_{-\lambda}(z)-I_\lambda(z)\right)}\nonumber\\
  &=&\frac12\Gamma(\lambda)\Gamma(1-\lambda)\Bigg(
  \frac{(z/2)^{-\lambda}}{\Gamma(1-\lambda)}\left(1+\frac{(z/2)^2}{1-\lambda}
  +\ldots\right)-\frac{(z/2)^\lambda}{\Gamma(1+\lambda)}\left(1
  +\frac{(z/2)^2}{1+\lambda}+\ldots\right)\Bigg)\nonumber\\
  &=&\frac12\Gamma(\lambda)\pfrac{z}2^{-\lambda}\left(1
  +\frac{(z/2)^2}{(1-\lambda)}+\ldots\right)+\frac12\Gamma(-\lambda)
  \pfrac{z}2^\lambda\left(1+\frac{(z/2)^2}{(1+\lambda)}+\ldots\right),
  \nonumber\\
\end{eqnarray}
where we have used
\begin{eqnarray}
\lefteqn{I_\lambda(z)\ =\ \pfrac{z}2^\lambda\sum_{k=0}^\infty
  \frac{(z^2/4)^k}{k!\Gamma(\lambda+k+1)}}\nonumber\\
  &=&\frac{(z/2)^\lambda}{\Gamma(1+\lambda)}
  \Bigg(1+\frac{(z/2)^2}{(1+\lambda)}+\frac{(z/2)^4}{2(1+\lambda)(2+\lambda)}
  +\frac{(z/2)^6}{6(1+\lambda)(2+\lambda)(3+\lambda)}+O(z^8)\Bigg).
  \nonumber\\
\end{eqnarray}

\end{appendix}

\end{document}